# Generation of relativistic polarized electron beams via collective beam-target interactions

**Authors:**
Xing-Long Zhu[1*], Min Chen[2,3], Wei-Min Wang[4,3], and Zheng-Ming Sheng[2,3,5*]

**Affiliations:**

[1] Institute for Fusion Theory and Simulation, School of Physics, Zhejiang University, Hangzhou 310058, China

[2] Key Laboratory for Laser Plasmas (MOE), School of Physics and Astronomy, Shanghai Jiao Tong University, Shanghai 200240, China

[3] Collaborative Innovation Center of IFSA, Shanghai Jiao Tong University, Shanghai 200240, China

[4] Department of Physics and Beijing Key Laboratory of Opto-electronic Functional Materials and Micro-nano Devices, Renmin University of China, Beijing 100872, China

[5] Tsung-Dao Lee Institute, Shanghai Jiao Tong University, Shanghai 201210, China

[*]Email: xinglong.zhu@zju.edu.cn (X.L.Z.); zmsheng@sjtu.edu.cn (Z.M.S.)

**Abstract:** Relativistic polarized electron beams can find applications in broad areas of fundamental physics. Here, we propose for the first time that electron spin polarization can be realized efficiently via collective beam-target interactions. When a relativistic unpolarized electron beam is incident onto the surface of a solid target with a grazing angle, strong magnetic fields are induced at the target surface due to the formation of a high reflux of target electrons. This results in violent beam self-focusing and corresponding beam density increase via magnetic pinching. The pinched dense beam in turn further enhances the magnetic fields to the level of a few Giga-Gauss, which is high enough to trigger strong synchrotron radiation of ultrarelativistic electrons. During the interaction, electron spin polarization develops along the magnetic field direction, which is achieved via radiative spin flips in the quantum radiation-dominated regime. As a consequence, the incident electron beam can be effectively polarized via the spin-dependent radiation reaction, for example, the mean polarization of electrons with energy less than 2 GeV can reach above 50% for an initial 5GeV beam. This provides a robust way for the development of polarized electron sources.

Polarized electrons play a crucial role in fundamental and applied research, as they allow one to study spin dependent or parity violation related physics, such as understanding the magnetic properties of materials [1], probing nuclear structures [2-4], and exploring new physics beyond the standard model [5]. Moreover, polarized electron beams can also be used as seed sources to produce polarized photon beams and polarized positron beams [6-9], which can be used to explore some fundamental problems such as the asymmetry between matter and antimatter in the universe. At present, high-energy polarized electrons are mainly generated in storage rings via radiative spin flips or spin polarization effect (so-called Sokolov-Ternov effect) [10-12], or extracted directly from photocathodes [13-15] and then accelerated in accelerators. This method typically requires large-scale accelerators and has been realized with limited currents. Since relativistic electrons undergo the Sokolov-Ternov effect in strong magnetic fields, it has recently been proposed that the strong magnetic fields of an ultra-intense laser may be used to polarize electrons [16-19]. However, it is difficult to obtain a highly polarized electron beam due to the spatiotemporal oscillation characteristics of the laser fields [20, 21]. To address this problem, several concepts have been proposed to break the symmetry of the laser fields for producing highly polarized electrons, such as colliding an electron beam with an asymmetric laser field [22-24], or trapping electrons at the magnetic nodes of a standing wave formed by two colliding circularly polarized laser pulses [25, 26], but they typically require very precise beam alignment and high intensity optical fields at the 10PW class. Currently, experimental realization of these ideas is technically very demanding both for the special laser fields and interaction configurations required. Alternatively, it was put forward to generate polarized electrons based upon plasma wakefield accelerators [27-31], which usually requires pre-polarized plasma or the injection of spin-polarized electrons. However, achieving such pure pre-polarized plasma sources remains challenging. Moreover, it requires the synergy of multiple laser pulses and proper control of the depolarization of electrons during trapping and acceleration. Simple and efficient ways to generate highly polarized electron beams are highly desired.

In this Letter, we present a scheme to achieve effective electron spin polarization simply via direct beam-solid interactions, where a relativistic unpolarized electron beam is used to impinge the surface of a solid target with a grazing angle. It is found that strong quasi-static magnetic fields are induced at the surface, which can act as the magnetic beam pinch and polarizer, focusing and polarizing the electron beam simultaneously, as shown schematically in Fig. 1. With spin-resolved particle-in-cell (PIC) simulations, we demonstrate the electron polarization process and illustrate the synergistic interplay between beam self-focusing and magnetic field-induced polarization. As a result, beam polarization can be attained with a simple configuration of direct beam-solid interactions without the use of ultra-intense laser pulses. This not only reveals a novel mechanism for generating highly polarized electrons, but also could be implemented experimentally as the beam parameters are likely to be available in the near future.

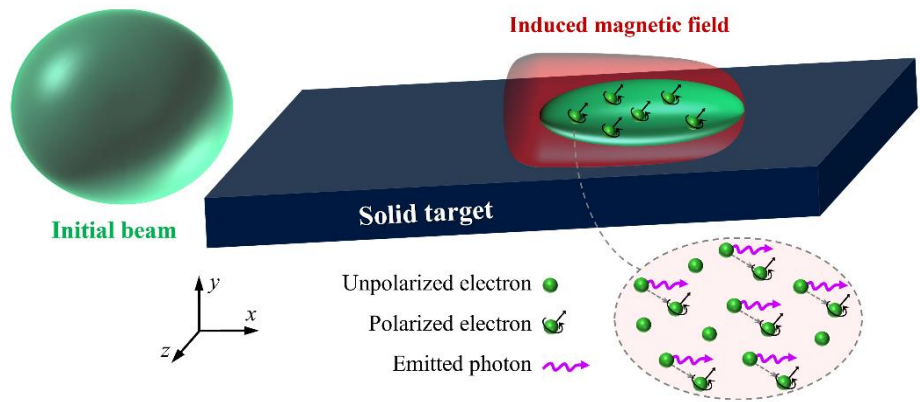

**Fig. 1.** Schematic diagram of electron spin polarization in beam-target interactions. An ultra-relativistic unpolarized electron beam is incident onto the surface of a solid target at grazing incidence, generating a super-strong magnetic field at the Giga-Gauss level, where the electron beam can be polarized via the radiative spin effect.

Physically, when the high-current electron beam passes through the plasma target, a large return current of plasma electrons moving in the opposite direction is rapidly established to maintain the current neutralization, which induces an intense asymmetric magnetic field at the target surface. This self-generated magnetic field creates a strong pinching force on the electron beam, causing the beam to focus, which in turn strengthens the magnetic field. In the meanwhile, the field serves as a magnetic polarizer that polarizes electrons via radiative spin flips. We have carried out three-dimensional (3D) simulations using the spin-resolved PIC code KLAPS [32, 33] to investigate beam-plasma interactions, where photon emission and spin polarization modules have been incorporated and have been fully benchmarked. Notice that the PIC method can be used to self-consistently simulate high-field beam-plasma interactions, as shown in the literature [34-40]. As an example, we adopt a 5GeV electron beam to strike a solid carbon target surface at grazing incidence, where both beam and target parameters are tunable. The target inclination angle is defined as $\theta_i = \mathrm{atan}(d_y/L_x) \times 180°/\pi \approx 1.1°$, where $L_x = 100 \mu m$ is the longitudinal length of the target along the $x$ direction and $d_y = 2\mu m$ is its transverse length along the y direction. The electron beam moves along the $x$-axis with a charge of about 2.7nC, which is characterized with an energy spread of 5%, a normalized emittance of 4 mm-mrad, and Gaussian spatial distributions both in the transverse and longitudinal directions with the dimensions of $\sigma_\perp = 2\mu m$ and $\sigma_\parallel = 1\mu m$. Comparable beam parameters are achievable with current accelerator techniques. For example, laser plasma accelerators are ideal for generating such high-current electron beams [41, 42], and recent advances have demonstrated that high-current nC-class electron beams with energies up to 10 GeV can be produced [43]. They can also be achieved with advanced accelerators [44, 45]. For instance, the E-332 experiment at the advanced stage of FACET-II following the proposal [36] has such parameters, where the beam has about $10^{21}$ cm$^{-3}$ density, 2nC charge, 1μm-scale size, and 10 GeV energy. Other synergic experiments such as E-308 alone can achieve a beam density of $10^{21}$ cm$^{-3}$ [46]. The maximum strength of the beam self-fields can reach about $10^{13}$ V/m, which is strong enough to ionize the target into plasma. Thus, it is reasonable to use a pre-ionized carbon plasma. For field ionization and other effects in our simulation, one may refer [47]. The size of the simulation window is $5\mu m \times 10\mu m \times 10\mu m$ in $x \times y \times z$ directions with grid cells of $250 \times 500 \times 500$, where 16 macroparticles per cell for each particle species are employed, and the absorption boundary condition is used. We employ the moving window technique to track the evolution of beam-target interactions and save the computing resources.

The key features of the electron-beam-solid interactions are illustrated in Fig. 2. For comparison, we consider the beam-target interaction in two different configurations, i.e., the electron beam impinges on the surface of the solid target at grazing incidence for case 1 [see Figs. 2(a-c)] and the electron beam hits the solid target with normal incidence for case 2 [see Figs. 2(d-f)]. The insets in Fig. 2(a) and Fig. 2(d) represent schematic diagrams of case 1 and case 2, respectively. Except for the interaction configuration, the other parameters are the same for both cases. It is well known that electron beam self-focusing does not occur when an electron beam interacts with a solid target, similar to case 2. Due to the current filamentation instability, a strong magnetic field can be induced to increase the local beam density. Nevertheless, the generated magnetic field is irregular as shown in Fig. 2(f), making it difficult to polarize the electron beam.

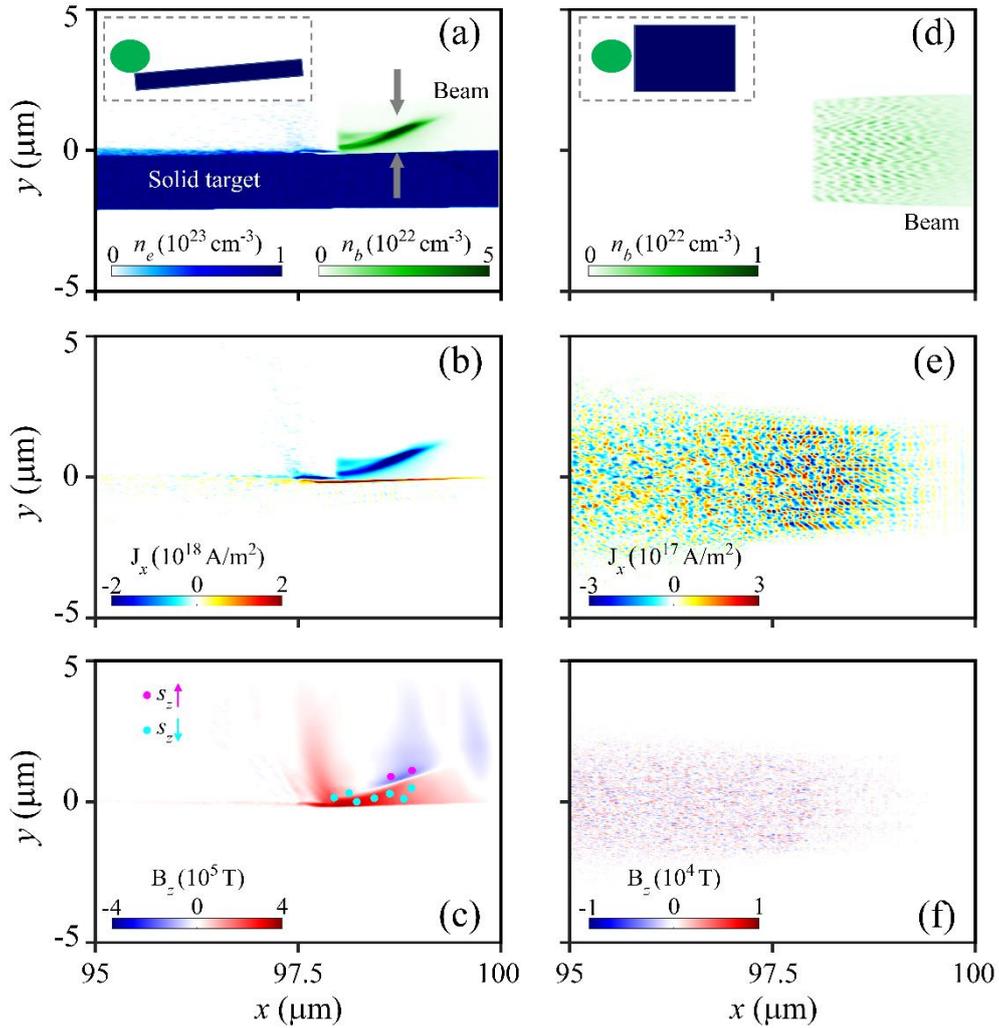

**Fig. 2.** Comparison of beam-solid interaction in two cases: in case 1, (a-c) the electron beam is incident on the surface of the target with grazing incidence, and in case 2, (d-f) the electron beam interacts with the target at normal incidence. The insets in (a) and (d) show the corresponding schematic diagrams. The spatial distributions in the $(x, y)$ plane of (a, d) the beam density ($n_b$), (b, e) the longitudinal electric current density ($J_x$), and (c, f) the magnetic field ($B_z$).

When a relativistic high-current electron beam hits the surface of a solid target with grazing incidence, it causes a large backflow of target electrons, and thus forms a strong asymmetric surface magnetic field, as shown in case 1. Under the interaction of such a magnetic field, the electron beam will undergo magnetic pinch and thus be focused, which in turn further enhances the field to a level of up to 4 Giga-Gauss. Consequently, the magnetic field generated in case 1 is more than two orders of magnitude higher than that generated in case 2, as illustrated in Figs. 2(c) and 2(f). More importantly, the beam can be polarized via the radiative spin flip effect, where the spin vector of emitting electrons after photon emission tends to be antiparallel to the magnetic field direction. This can be thought of as a miniature magnetic polarizer. Since the magnetic field generated on the surface of the solid target is stronger, it is more likely to trigger photon emission in the strong field region, so that the emitting electrons can obtain net spin polarization. Finally, an extremely dense beam of polarized electrons can be generated, which is difficult to realize with other methods. Compared with previously proposed methods based on laser-electron collisions [22-26], this method has the unique advantage of achieving extreme magnetic field generation and spin polarization through the direct interaction of an electron beam with a solid target, without the use of high-intensity lasers, which is robust and simple.

In the configuration of grazing incidence, the electron beam is strongly focused due to the generation of the strong surface magnetic field. A similar focusing phenomenon was first observed in our previous work in order to efficiently produce bright gamma-rays [37], but its potential for producing spin-polarized electrons has not been revealed. Actually, the generation of strong magnetic fields does not lead naturally to the production of a polarized electron beam [47]. Here we identify a parameter regime and the involved physics for producing highly polarized dense electrons with this unique interaction configuration. Figure 3(a) shows the energy-position trajectories of representative electrons focused along the target surface. The induced effective field $\mathbf{E}_{\text{eff}} = \mathbf{E}_\perp + \boldsymbol{\beta} \times \mathbf{B}$ increases with the beam focusing, and its amplitude is comparable to the electron beam self-field that can be scaled as $E_{\text{eff}} \sim E_b \propto 4\pi e n_b \sigma_\perp^2 / r$, where $e$ is the elementary charge, $n_b$ is the beam density, $r$ is the focused beam radius, $\mathbf{E}_\perp$ is the transverse electric field, $\mathbf{B}$ is the magnetic field, $\boldsymbol{\beta} = \mathbf{v}/c$ is the normalized electron velocity, and $c$ is the speed of light in vacuum. The importance of QED effects can be characterized by $\chi_e =$ $\gamma_e|\mathbf{E}_{\text{eff}}|/E_s$, where $E_s = 1.3 \times 10^{18}$ V/m is the Schwinger field [48]. As the beam focusing develops, a stronger surface field can be induced, allowing the QED effects to be efficiently triggered. A large amount of electron energy can be converted into high-energy photon emission [see Fig. 3(a) for the electron energy evolution], leading to an increase in the number of low-energy electrons after radiation, as shown in Fig. 3(b). In the meanwhile, the electrons emitting photons can obtain spin polarization via the radiative spin flip effect. For example, if we consider these electrons with energies less than 2GeV, the value of their average polarization can reach $|S_z| \approx 70\%$. As the field increases, the beam electrons away from the target surface will also be polarized, but the resulting spin state is opposite to the spin state of the emitting electrons near the target surface, resulting in the total spin polarization value decreasing to 51%. When counting highly polarized electrons with energy <2GeV after photon emission, they have a charge of about 25pC, and normalized emittances of about 10 mm-mrad in the z direction and 22 mm-mrad in the y direction.

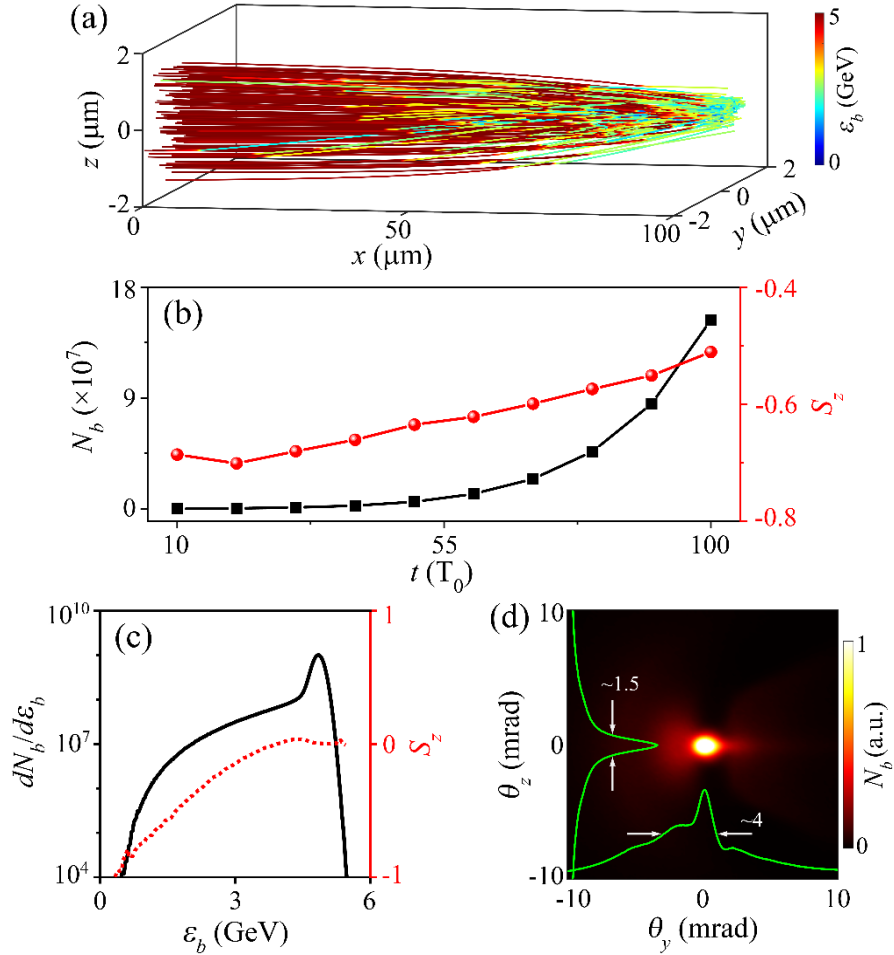

**Fig. 3.** (a) Evolution of electron trajectories and corresponding energies in 3D geometry during the beam-solid interaction

with grazing incidence. (b) The total number $N_b$ and the average spin polarization $S_z$ of the emitting electrons with energies <2GeV after photon emission as a function of the interaction time $t$, where $T_0 = 3.33$fs. (c) Distributions of the electron number and average spin polarization versus the energy $\varepsilon_b$. (d) The angular distribution of the resulting electron beam, where the green lines indicate the divergence angles along $\theta_y$ and $\theta_z$, respectively.

In order to describe the spin polarization process due to photon emission, we adopt a fully spin-resolved Monte Carlo method [32, 33, 49] as follows

$$\frac{d^2 W_{\text{rad}}}{dudt} = C_{\text{rad}}(w_{\text{rad}} + \boldsymbol{g}_{\text{rad}} \cdot \boldsymbol{S}_i + \boldsymbol{p}_f \cdot \boldsymbol{S}_f + \boldsymbol{p}_\xi \cdot \boldsymbol{\xi})/4, \tag{1}$$

where $C_{\text{rad}} = \alpha m_e^2 c^4/\sqrt{3}\pi\hbar\varepsilon_e$, $w_{\text{rad}} = \left(\frac{u^2-2u+2}{1-u}\right) K_{2/3}(y_1) - \text{Int}K_{1/3}(y_1)$, $\alpha \approx 1/137$ is the fine structure constant, $m_e$ is the electron mass, $\hbar$ is the reduced Planck constant, $K_n(y)$ is the $n$-order modified Bessel function of the second kind, $\text{Int}K_{1/3}(y) \equiv \int_y^\infty K_{1/3}(x)dx$, $y_1 = 2u/[3(1-u)\chi_e]$, $u = \varepsilon_\gamma/\varepsilon_e$, $\varepsilon_e$ is the emitting electron energy, $\varepsilon_\gamma$ is the emitted photon energy, $\boldsymbol{g}_{\text{rad}} = -uK_{1/3}(y_1)\boldsymbol{e}_2$, $\boldsymbol{p}_f = -\frac{u}{1-u}K_{1/3}(y_1)\boldsymbol{e}_2 + [2K_{2/3}(y_1) - \text{Int}K_{1/3}(y_1)]\boldsymbol{S}_i + \frac{u^2}{1-u}[K_{2/3}(y_1) - \text{Int}K_{1/3}(y_1)](\boldsymbol{S}_i \cdot \boldsymbol{e}_v)\boldsymbol{e}_v$, and $\boldsymbol{p}_\xi = \frac{u}{1-u}K_{1/3}(y_1)(\boldsymbol{S}_i \cdot \boldsymbol{e}_1)(1,0,0) + \left[\frac{2u-u^2}{1-u}K_{2/3}(y_1) - u\text{Int}K_{1/3}(y_1)\right](\boldsymbol{S}_i \cdot \boldsymbol{e}_v)(0,1,0) + \left[K_{2/3}(y_1) - \frac{u}{1-u}K_{1/3}(y_1)(\boldsymbol{S}_i \cdot \boldsymbol{e}_2)\right](0,0,1)$. Here $\boldsymbol{S}_i$ and $\boldsymbol{S}_f$ are respectively the initial and final spin vectors of the electrons due to photon emission, $\boldsymbol{e}_v$ is the unit vector along the electron velocity, $\boldsymbol{e}_1$ is the unit vector along the transverse acceleration of electrons, and $\boldsymbol{e}_2 = \boldsymbol{e}_v \times \boldsymbol{e}_1$. The photon polarization can be characterized by Stokes parameters $\boldsymbol{\xi} = (\xi_1, \xi_2, \xi_3)$ with $|\boldsymbol{\xi}| = 1$, defined with respect to the orthonormal basis vector $(\boldsymbol{e}_1, \boldsymbol{e}_2, \boldsymbol{e}_v)$. In the interaction configuration, the induced surface electric and magnetic fields are nearly perpendicular to the driving electron velocity. Considering that the emitting electrons and radiated photons are mainly directed in the $x$-$y$ plane, the magnetic field direction in their respective rest frames can be approximately directed along $\boldsymbol{\zeta} \approx (0,0,B_z/|B_z|)$. For simplicity, the spin vector of electrons is defined with respect to $-\boldsymbol{e}_2$, where $-\boldsymbol{e}_2 \approx -\boldsymbol{\zeta}$, indicating that the spin vector of electrons after high-energy photon emission tends to be antiparallel to the magnetic field direction. Hence, Eq. (1) can be further reduced to $\frac{d^2 W_{\text{rad}}}{dudt} = \frac{C_{\text{rad}}}{4}\{w_{\text{rad}} + uK_{1/3}(y_1)S_i + \frac{u}{1-u}K_{1/3}(y_1)S_f + [2K_{2/3}(y_1) - \text{Int}K_{1/3}(y_1)]S_i S_f + [K_{2/3}(y_1) + \frac{u}{1-u}K_{1/3}(y_1)S_i]\xi_3\}$. Accordingly, the averaged polarization of the emitting electron after photon emission can be described as

$$\bar{S}_f = \frac{\left(\frac{u}{1-u}\right)K_{1/3}(y_1) + [2K_{2/3}(y_1) - \text{Int}K_{1/3}(y_1)]S_i}{w_{\text{rad}} + uK_{1/3}(y_1)S_i}, \qquad (2)$$

which reveals the distinct characteristics of spin polarization between low-energy and high-energy states.

We now elucidate the underlying physics as follows. When emitting low-energy photons with $\varepsilon_\gamma \ll \varepsilon_e$, the electron can keep its polarization almost unchanged with $\bar{S}_f \to S_i$, that is, low energy photon emission has little effect on the spin polarization process. Since the driving electron beam is initially unpolarized, the high-energy electrons have relatively low spin polarization after emitting low-energy photons. While for emitting high-energy photons with $\varepsilon_\gamma \to \varepsilon_e$, where the electron spin vector tends to be antiparallel to the magnetic field direction, such that the low-energy electrons have relatively high polarization after emitting high-energy photons. This is in good agreement with the simulation results shown in Fig. 3(c). For example, these electrons after emitting low-energy photons have a low polarization value $|S_z| \to 0$ at energies of about 5GeV (where $\varepsilon_\gamma \ll \varepsilon_e$), while these electrons after emitting high-energy photons have a high polarization value $|S_z| \to 1$ at energies of <1GeV (where $\varepsilon_\gamma \to \varepsilon_e$). During the interaction, the electron beam can be well confined along the target surface due to magnetic pinch and is always kept within a divergence angle of a few mrad in the transverse directions, as illustrated in Fig. 3(d).

In order to demonstrate the robustness of the scheme, we investigate the effects of the driving electron beam energy and the interaction target length on the spin polarization process, while keeping other parameters unaltered. It is indicated that the scheme can be applied to the driving beams with different energies, as shown in Fig. 4(a). As explained earlier, these parent electrons can keep their spin polarization almost unchanged (that is $|S_f| \to |S_i| \sim 0$) when emitting low energy photons with $\varepsilon_\gamma \ll \varepsilon_e^i$, where $\varepsilon_e^f \to \varepsilon_e^i$, $\varepsilon_e^f$ and $\varepsilon_e^i$ are respectively the initial and final energies of the electrons after photon emission; While for emitting high-energy photons with $\varepsilon_\gamma \to \varepsilon_e^i$, the spin state of the electron after photon emission tends to be antiparallel to the magnetic field direction via the radiative spin flip effect, and a large amount of energy is lost into high-energy photon emission, where $\varepsilon_e^f \ll \varepsilon_e^i$, leading to higher spin polarization (that is $|S_f| \to 1$).

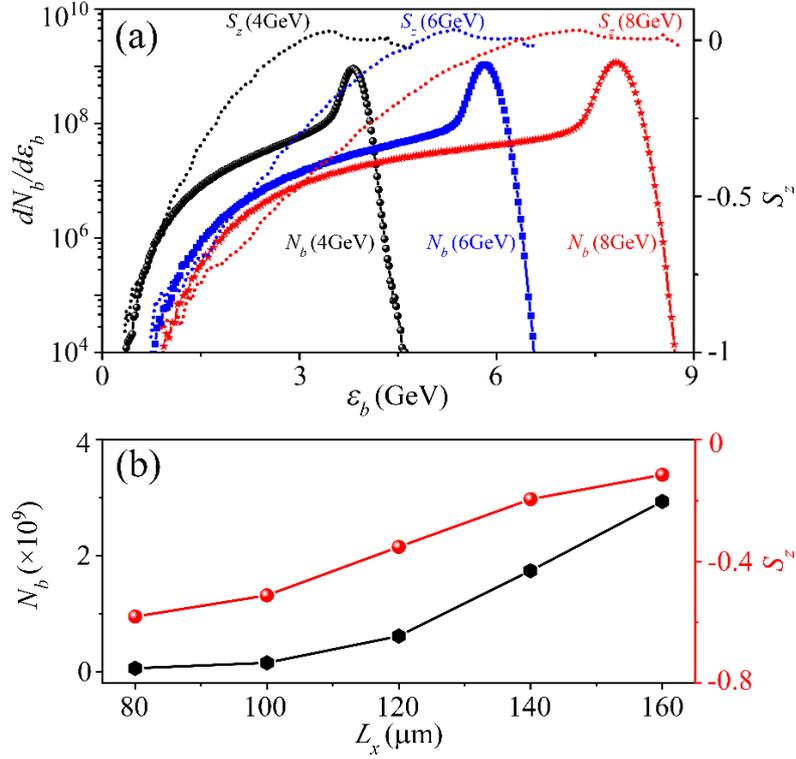

**Fig. 4.** (a) The energy spectrum and spin polarization of the electron beam with different initial energies: 4GeV (black color), 6GeV (blue color), and 8GeV (red color). (b) The total number (black line) and the average spin polarization (red line) of electrons with energies <2GeV after high-energy photon emission as a function of the longitudinal length of the solid target, where the initial beam energy is 5 GeV.

In Fig. 4(b), we illustrate the dependence of the mean spin polarization $S_z$ of electrons with energies <2GeV and their number $N_b$ on the interaction target length $L_x$. This is actually equivalent to the effect of the target inclination angle $\theta_i$, because $\theta_i = \mathrm{atan}(d_y/L_x) \times 180°/\pi$ varies with $L_x$ for a given $d_y = 2\mu m$. In other words, $L_x$ changing from 80μm to 160μm is equivalent to $\theta_i$ changing from 1.43° to 0.72°. It is found that an appropriately long interaction target is conducive to beam focusing and thus produces a stronger surface magnetic field. This significantly enhances photon emission, causing an increase in the number of low-energy electrons after the emission of copious high-energy photons. As the surface magnetic field increases, the beam electrons away from the target surface will be polarized by a magnetic field of opposite polarity, producing spin polarization opposite to the spin polarization of beam electrons near the target surface. As a whole, the overall spin polarization of the beam will be reduced. For example, when $L_x = 80\mu m$, the total number of electrons with energies <2GeV after photon emission is about $6 \times 10^7$ with a spin polarization of about $|S_z|\sim 0.6$; When $L_x = 160\mu m$, the number of emitting electrons increases to $2.9 \times 10^9$ while their spin polarization decreases to $|S_z|\sim 0.12$. Therefore, the average spin polarization and the total number of polarized electrons can be tuned by simply changing the target parameters for different application requirements. Since the spin polarization of energetic electrons depends on their energy, a spectrometer beamline [50] can be used to capture polarized electrons, perform selection of specific energies, and deliver them for applications.

It should be noted that the present work for polarized electron generation is different from our previous studies on polarized positron generation [32] as they are in different parameter regimes and with different interaction configurations. The present scheme works in the low QED parameter regime and there is almost no positron produced. In the previous work on polarized positron production, the driving electron beam is first pre-focused through a hollow cone target to a density much higher than that in the present scheme, and then strikes the surface of a solid target with a normal incidence, triggering efficient production of copious pairs and intense gamma-rays with high QED parameters. As the electron beam experiences both positive and negative magnetic fields during extreme magnetic pinching, the electron beam cannot be well polarized, see [47] for more details. Meanwhile, the positrons produced via the multiphoton Breit-Wheeler process are mainly located inside the target. Therefore, they experience unipolar strong magnetic fields and can be well polarized via radiative spin flips.

In conclusion, we have discovered a new regime and configuration of beam-solid interactions suitable for the generation of polarized dense electron beam. When a relativistic electron beam interacts with the surface of a solid target at grazing incidence, strong asymmetric magnetic fields are generated by return background electrons in the target. The electron beam then undergoes strong self-focusing due to magnetic pinch in such magnetic fields, which further enhances the field to the Giga-Gauss level, high enough to act as a spin polarizer through the radiative spin flip effect. Since the spin vector of the emitting electrons after photon emission tends to be antiparallel to the magnetic field direction, the beam can be naturally polarized in such a strong asymmetric surface field. These processes develop naturally and do not require additional devices or high-intensity laser fields. We have demonstrated the above beam-solid interaction processes and the spin polarization dynamics using spin-resolved PIC simulations. The results could be interesting for cutting-edge applications in nuclear and particle physics [1-5] and high field physics [51, 52], etc.

## Acknowledgements

This work was supported by the National Natural Science Foundation of China (Grant Nos. 12205186 and 12135009).